# New direction and perspectives in elastic instability and turbulence in various viscoelastic flow geometries without inertia

(Short Review)

Victor Steinberg

*Weizmann Institute of Science, Rehovot, Israel*
E-mail: victor.steinberg@weizmann.ac.il

We shortly describe the main results on elastically driven instabilities and elastic turbulence in viscoelastic inertialess flows with curved streamlines. Then we describe a theory of elastic turbulence and prediction of elastic waves Re $\ll$ 1 and Wi $\gg$ 1, which speed depends on the elastic stress similar to the Alfven waves in magneto-hydrodynamics and in a contrast to all other, which speed depends on medium elasticity. Since the established and testified mechanism of elastic instability of viscoelastic flows with curvilinear streamlines becomes ineffective at zero curvature, so parallel shear flows are proved linearly stable, similar to Newtonian parallel shear flows. However, the linear stability of parallel shear flows does not imply their global stability. Here we switch to the main subject, namely a recent development in inertia-less parallel shear channel flow of polymer solutions. In such flow, we discover an elastically driven instability, elastic turbulence, elastic waves, and drag reduction down to relaminarization that contradict to the linear stability prediction. In this regard, we discuss shortly normal versus non-normal bifurcations in such flows, flow resistance, velocity and pressure fluctuations, and spatial and spectral velocity as function of Wi at high elasticity number.

Keywords: elastic instability, elastic turbulence, viscoelastic inertialess flows.

## 1. Introduction

An addition of the tiny amount of long flexible linear polymer molecules into viscous Newtonian solvent strongly affects a laminar flow caused by flow stretched polymers [1–3]. Even in low-velocity flows, discussed here, significant elastic stress at large velocity gradients and high polymer relaxation time could be generated. Indeed, in laminar shear flows with curved streamlines, elastic instabilities and elastic turbulence (ET) occur at a low Reynolds number, Re $\ll$ 1, and high Weissenberg number, Wi $\gg$ 1, corresponding to the large fluid elasticity number, El = Wi/Re $\gg$ 1 [4]. Here the control parameters are Wi = $\lambda U/L$ and Re = $\rho UL/\eta$, where $U$ is the mean flow velocity, $L$ is the vessel size, $\rho$ and $\eta$ are the fluid density and dynamic viscosity, respectively, and $\lambda$ is the longest polymer relaxation time [4]. Elastic stress in a viscoelastic fluid destabilizes an inertialess flow with curvilinear streamlines at Wi $\geq$ Wi$_c$ and further at Wi $\gg$ Wi$_c$ leads to ET, where Wi$_c$ is the Weissenberg number of the instability onset. In curvilinear shear flows, an elastic instability is driven by the first normal stress difference along the curved streamlines $N_1$, the "hoop stress", which initiates a bulk force in the direction of curvature giving rise to the instability [1–3]. This mechanism, however, becomes ineffective in parallel shear flows with zero curvature streamlines [1–3, 5]. In this review, first, I shortly describe the elastic instability and ET in polymeric solution flows with the curved streamlines their characterization and main properties, such as dependencies of the friction factor and rms velocity and pressure fluctuations on Wi, scaling decay of velocity and pressure power spectra, strong enhancement of mixing, and particle pair dispersion in ET. These results are already observed and partially explained in numerous flow geometries with curved streamlines during the last about two decades (see, e.g., Refs. 6 and 7).

However, the main subject of this review is to discuss recent experimental progress, accompanied by numerous unexpected discoveries and findings, elastic instability and ET in a straight channel inertiales flow of a viscoelastic fluid. In spite of the ineffectiveness of the hoop stress mechanism and mathematically proved linear stability of parallel shear flow geometries [8, 9], it does not imply their global stability. This progress is similar to the development that took place for the last two–three decades in extensive studies of parallel shear flows of Newtonian fluids [10].

Indeed, the experiments reveal first in a pipe [11] and later in a square microchannel [12–14] of viscoelastic flows with strong prearranged velocity perturbations at the inlet that an elastic instability takes place and results in a direct transition to a chaotic flow just at Wi > Wi$_c$. Further progress in a straight planar channel flow of a viscoelastic fluid with and without prearranged strong disturbances, occurred during the last several years, is the main subject of this review.

## 2. Elastic instability and ET in polymer fluid flows with curved streamlines

### 2.1. Pure elastic linear instabilities in flows with curved streamlines

As mentioned in Introduction, the theoretical description and physical explanation of the mechanism of the elastically driven instability are suggested more than thirty years ago to explain the Taylor–Couette viscoelastic inertia-less flow between two cylinders [1]. Later on, the elastic instability mechanism is generalized by using scaling approach for all flow geometries with curvilinear streamlines [5]. During the years, numerous experimental, theoretical, and numerous studies of the elastically driven instability of numerous flow geometries with curved streamlines of viscoelastic fluids verify all aspects of the instability criterion (see, e.g., summary in [7]). Thus, in such flows, the elastic instability occurs solely due to elastic stress, $\sigma$, the only source of nonlinearity at Re ≪ 1, generated by polymers stretched in a laminar shear flow. In shear flows, the elastic stress is anisotropic and characterized by the first normal stress difference, $N_1 = \sigma_{11} - \sigma_{22}$ in 2D velocity plane, where $\sigma_{11}$ and $\sigma_{22}$ are the diagonal components of the stress tensor, $\sigma_{ij}$, along the azimuthal and radial flow directions, respectively [4]. Since $|\sigma_{11}| \gg |\sigma_{22}|$ and so $N_1 \approx \sigma_{11}$, the azimuthal elastic stress component, "hoop stress", generates a bulk force acting on the polymer solution towards the center of curvature, leading to the striking rod climbing, or Weissenberg, effect, discovered and described long time ago [15]. The same bulk force triggers an elastic instability in shear flows with curved streamlines [1–3, 5–7]. The ratio of the elastic stress per unit volume and time, $(U\nabla)\sigma$, to its dissipation due to relaxation of the elastic stress per unit volume, $\sigma/\lambda$, defines the instability onset. Indeed, the main control parameter, the Weissenberg number, Wi ~ $(U\nabla)\sigma/(\sigma/\lambda)$ ~ $\lambda U/L = \lambda/\tau_u$, where $\tau_u = L/U$ is the hydrodynamic time scale, and the instability onset takes place at Wi > 1 and Re ≪ 1. We would like to emphasized that elastic instabilities found in all shear flows with the curvilinear streamlines are linear normal mode bifurcations either forward (continuous) or backward (hysteretic). It means that above the instability onset only fastest growing mode shows up and saturates at sufficiently large amplitude. In the case of the forward bifurcation, the transition is characterized by the power-law dependence of the main parameters on the distance from the onset, Wi – Wi$_c$, with the exponent 0.5, as followed from the corresponding Landau equation [16]. An example of such an elastic instability is a creeping viscoelastic flow between two obstacles hindering a channel flow, studied and characterized in details in [17] as well as a flow past an obstacle in a channel flow [18, 19]. In spite of the proved linear stability of a channel shear flow [8, 9], the linear elastic instabilities of downstream as well as upstream wakes and flow between two widely spaced obstacles hindering the channel flow are found in all of them [17–19] and are described in Chapter 3.

### 2.2. Elastic turbulence and its main properties

The discovery of the ET main features and its quantitative characterization suggests an analogy of a chaotic flow of dilute polymer solution at Re ≪ 1 and inertial turbulence at Re ≫ 1. The experimental setup [20] consists of a stationary cylindrical cup with a plain bottom (the lower stationary disk; radius 43.6 mm), concentric with a rotating upper disk (radius 38 mm) attached to the shaft of a commercial rheometer, which just touches the surface of a solution. The solution consists of 65% saccharose and 1% NaCl in water with viscosity 0.324 Pas as a solvent for the polymer, to which polyacrylamide of $M_w$ = 18 MDa at a concentration of 80 ppm by weight has been added. The walls of the cup are transparent to allow Doppler velocimetry measurements. The lower disk is also transparent and, together with a mirror tilted by 45° placed under the lower disk, provides viewing access from below using a CCD camera mounted at the side. The temperature in the setup is stabilized by circulating air in a closed box and by circulating water at 12 °C under a steel lower plate. There are three main features, by which ET is distinguished [6, 20, 21]. (i) Large increase in the flow resistance is comparable with that of inertial turbulence at Re ≫ 1. (ii) Broad range of temporal scales with a steep power-law decay of the velocity power spectra at higher frequencies. And (iii) Orders-of-magnitude increase in a mixing efficiency compared with diffusion [6, 21, 22] [(Figs. 1(a)–(c)]. These features are first reported for a swirling flow between rotating and stationary disks as well as a serpentine channel flow [20, 21], and then, in other flow geometries during the years [6]. However, a formal similarity between ET and inertial turbulence does not imply the same underlying physical mechanism. Contrary to inertial turbulence driven by Reynolds stress with a local energy flux transfer in the inertial-scale range, ET is solely driven by the elastic stress without inertia. Moreover, the power-law decay of velocity power spectrum is not related to the cascade of the local energy transfer and conservation laws, since the leading mechanism of small-scale generation occurs via stretching and folding of the elastic stress field during random advection of a fluid element by the fluctuating velocity field at the largest scales. A back reaction of the elastic stress field on the flow leads to a dynamical stochastic stationary state of ET. Thus, the key experimental observation in ET is the power-law decay of the velocity power spectra in a frequency domain, $E_V(f) \sim f^{-\alpha}$. with the exponent α > 3, between 3.3 and 3.6 depending on the flow geometry [6, 20, 21, 23]. The velocity spectrum in the spatial domain is integrable due to its steep decay at small scales resulting in a sharp reduction of their contribution to both the velocity and velocity gradient fields. The latter is the signature of the velocity field smoothness since only large spatial scales of about vessel size define the statistics of the velocity field. Thus, ET is essentially a spatially smooth and temporally random flow, dominated by a strong nonlinear interaction of a few large-scale spatial modes. This type of random flow appears in inertial turbulence below the dissipation scale [24]. Later on [25, 26], it was found that frequency power spectra of both $E_{\Gamma\Gamma}$ of torque $\Gamma$, and $E_{pp}$ pressure $p$, fluctuations in the ET regime of a swirling flow also reveal algebraic decays with exponents μ ≈ 4 and β = 3 for $E_{\Gamma\Gamma} \sim f^{-\mu}$ and $E_{pp} \sim f^{-\beta}$, respectively [Figs. 1(d) and (e)]. Thus, there is another striking similarity with high-Re turbulence.

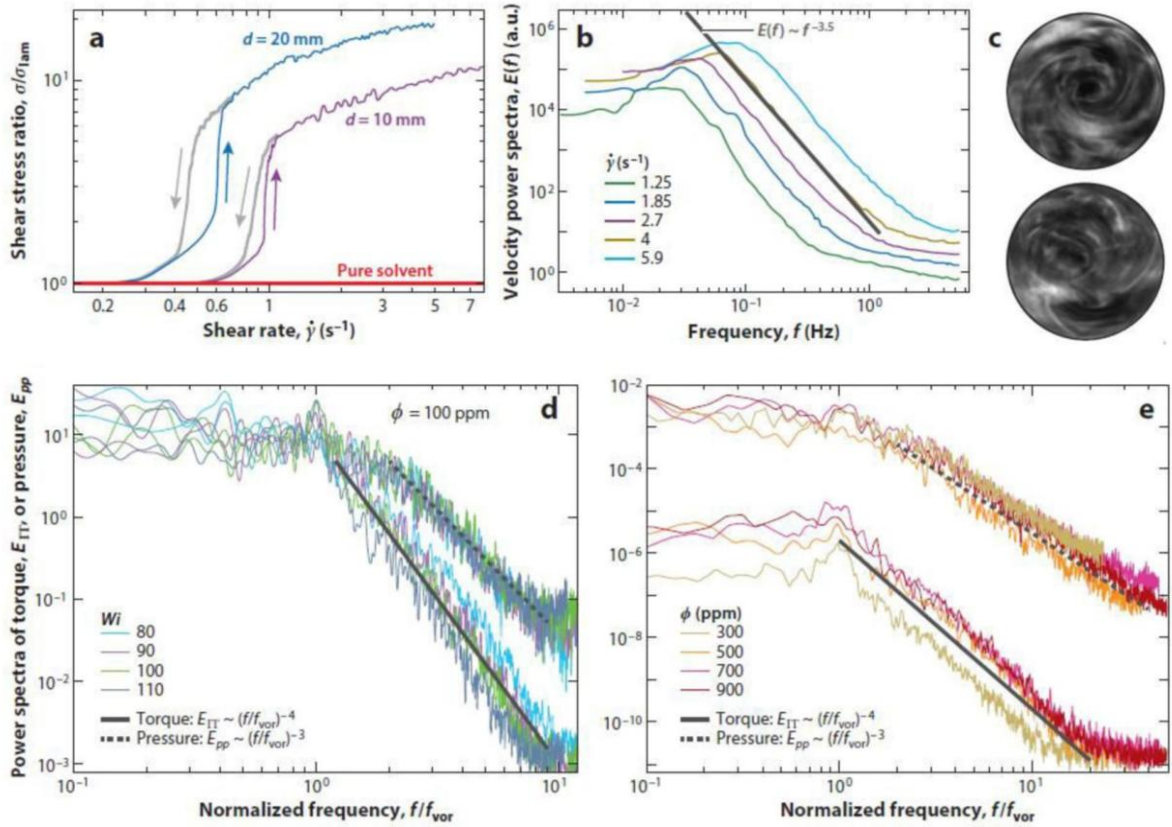

*Fig. 1.* (Color online) (a) The average stress ratio $\sigma/\sigma_{\text{lam}}$, or the flow resistance, versus the shear rate $\dot{\gamma}$ for two gap values between two disks, $d$ =10 mm, 20 mm, and a pure solvent. (b) Frequency power spectra of velocity fluctuations $E_V(f)$ at the center of the swirling flow for five $\dot{\gamma}$ above $\dot{\gamma}_c \approx 1$ s$^{-1}$. The fit of $E_V(f)$ decay at $\dot{\gamma} = 4$ s$^{-1}$ is $E_V(f) \sim f^{-3.5}$. (c) Two snapshots of irregular flow structures at Wi = 13 and Re = 0.7, visualized by suspended, light-reflecting flakes. (d) Power spectra of torque $\Gamma$, and pressure $p$, fluctuations $E_{\Gamma\Gamma}$ and $E_{pp}$ versus $f/f_{\text{vor}}$ [25]. (e) $E_{\Gamma\Gamma}$ and $E_{pp}$ versus $f/f_{\text{vor}}$ at maximum Wi. The value of the main vortex frequency $f_{\text{vor}}$ is obtained from the peaks in $E_{\Gamma\Gamma}$ and $E_{pp}$ versus Wi [25].

Finally, next remarkable feature of ET is a drastic enhancement of mixing efficiency visualized by fluorescent dye used as a passive tracer, quantitatively studied in a curvilinear channel flow for the first time (Fig. 2) [21]. It occurs in ET due to increased velocity and velocity gradient fluctuations [21, 22, 27]. For the same reason, strongly enhanced dispersion of micro-particles in a viscoelastic flow in both microfluidic devices and porous media, such as rock cores, is found [28, 29]. The latter occurs at Wi value, which coincides with the increase in oil displacement efficiency in 3D opaque porous structures. The latter has important industrial applications in oil recovery [29–32]. As found in [21, 22, 27, 33–35], a characteristic mixing time, proportional to a mixing length, decreases up to four orders of magnitude than that of molecular diffusion. Moreover, a more elaborate quantitative approach has revealed a key role of the velocity boundary layer near the wall in reducing mixing [33–35]. Due to reduced velocity near the wall, the boundary layer becomes a sink for passive tracers. The numerical simulations and experiment disclose that vigorous, random, and localized excursions of the tracers trapped near the walls strongly perturb the passive scalar distribution in the peripheral and bulk regions causing growth of the mixing length [33, 35]. To understand mixing and transport phenomena on a microscopic level, we study a particle pair dispersion that in ET is expected to show an exponential particle pair separation on average. However, the experiment in ET has revealed the ballistic pair dispersion for sufficiently long times [36] that is confirmed by numerical simulations later on [37]. The initial ballistic growth persists for extremely long times, above 200 the Kolmogorov timescale, before the exponential growth takes over [37].

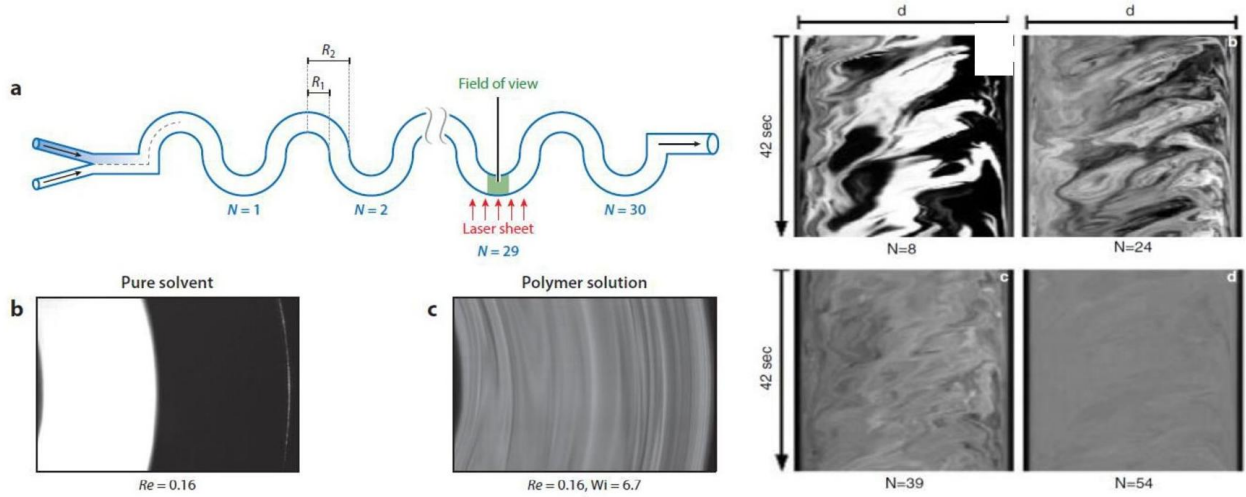

*Fig. 2.* (Color online) (a) Schematic of a curvilinear channel. (b), (c) Snapshots of flow mixing taken with a laser visualization sheet at location $N = 29$ channel, for (b) a pure solvent at Re = 0.16 and (c) a polymer solution at the same flow rate. (d) Space-time plots of flow mixing at different positions, $N$, along the channel. All flow-mixing images are taken at Wi = 6.7. Bright regions correspond to high concentrations of the fluorescent dye.

### 2.3. Theory of elastic turbulence

The theory of ET is based on the Navier–Stokes equation (NSE) for a polymer solution with a coupling term, engendered by the back reaction of the elastic stress field $\sigma_{ij}$ on the velocity field $V_i$, and the elastic stress dynamics equation for $\sigma_{ij}$ in the Oldroyd-B model [4]. In ET at Re << 1, an inertial term in NSE can be neglected. For sufficiently stretched polymers, one gets $\sigma_{ij} \gg \sigma^0_{ij}$, where $\sigma^0_{ij}$ is the elastic stress due to thermal noise, and two equations describing homogeneous and isotropic flow of a dilute polymer incompressible solution are the following [38]:

$$\nabla_i p = \eta s \Delta \otimes + \rho_F i, \quad (1)$$

$$\rho F_i = \eta_\rho \nabla_j \sigma_{ij}, \quad (2)$$

$$\nabla_i V_i = 0, \quad (3)$$

$$\partial_t \sigma_{ij} + (V_k \nabla_k)\sigma_{ij} = \sigma_{kj}\nabla_k V_i + \sigma_{ik}\nabla_k V_j - 2\sigma_{ij}/\lambda, \quad (4)$$

where $\eta$ and $\rho$ are the solution dynamic-viscosity and density, respectively. It is suggested in [38] that $\sigma_{ij}$ relaxes to statistically stationary state in the uniaxial form $\sigma_{ij} = B_i B_j$ at times much longer than the short correlation time of the velocity gradient. Indeed, in the ET regime, one finds $\sigma_{ij} \gg \sigma^0_{ij}$ and $R_i \gg R_0$, where $R_i$ is the end-to-end vector of a stretched polymer and $R_0$ is the gyration radius. Then instead of the expression $\sigma_{ij} \sim \langle R_i R_j \rangle$ averaged over thermal noise, one gets $\sigma_{ij} \sim R_i R_j$ in the uniaxial form due to neglecting $\sigma^0_{ij}$ [38]. Taking into account the uniaxial form $\sigma_{ij} = B_i B_j$, one can rewrite the set of equations for polymer hydrodynamics in ET (1)–(4) in the form similar to magneto-hydrodynamic (MHD) equations with the only difference between relaxation dissipation and magnetic resistivity [39]:

$$\nabla_i p = \beta \Delta V_i + (1-\beta)(B_j \nabla_j) B_i, \quad (5)$$

$$\nabla_i V_i = 0, \quad (6)$$

$$\partial_t B_i + (V_j \nabla_j) B_i = (B_j \nabla_j) V_i - 2B_i/\lambda, \quad (7)$$

$$\nabla_i B_i = 0, \quad (8)$$

where $\beta = \eta s/\eta$, $B_i$ is the solenoidal vector analogous to the director in nematics, in contrast to a magnetic field vector in MHD. At sufficiently large back reaction of the elastic stress $(B_j \nabla_j) B_i$, the flow becomes chaotic in a statistically stationary state defined as ET. The major result of the theory based on Eqs. (1)–(4) or (5)–(8) is the prediction of the power-law decay of the spherically normalized kinetic energy spectrum

$$E(k) \sim V^2 \rho l (kl)^{-\alpha} \text{ with } \alpha > 3. \quad (9)$$

Thus, in a contrast to inertial turbulence, where α has the well-defined value in the inertial range of scales due to the energy conservation law [16], in ET the range of the values of α is determined by the velocity smoothness condition [38].

The scaling expression in (9) follows from the theory based on two main assumptions. (i) A dynamical stochastically stationary state of ET occurs due to the back reaction of elastic stress on the initial flow. (ii) Viscous and relaxation dissipations are of the same order. The first means that the maximum Lyapunov exponent of ET (or $(\nabla_i V_j)_{rms}$), defining the polymer stretching rate, should be about $\lambda^{-1}$. And the second means that $\eta(\nabla_i V_j)^2 \sim B^2/\lambda$ [38]. Similarly one gets the scaling exponent of the elastic energy spectrum $E_B(k) \sim B^2 l(kl)^{-\nu}$, where $\nu = (\alpha - 2) > 1$, which is close to the passive scalar Batchelor decay exponent $\gamma = -1$ [24]. Here $k$ is the wave number, and $l$ is the length scale defining the largest average velocity gradient. From both the power spectrum expressions, one finds that the main energy in ET is carried out by the stretched polymers resulted in the elastic energy $\sim B^2 \gg \rho V^2$. Indeed, the latter follows from the relation $E(k) \sim \text{Re}(kl)^{-2} E_B(k)$, where $\rho(V/B)^2 \sim \text{Re} \ll 1$. Indeed, $B^2 \sim \eta/\lambda$ and $\rho V^2 \sim \rho(l/\lambda)^2$ leading to $\rho(V/B)^2 \sim \rho l^2/\eta\lambda = \text{El}^{-1} \sim \text{Re}$ [38], where $\rho$ is the fluid density, and the elasticity $\text{El} = \text{Wi}/\text{Re} = \eta\lambda/\rho l^2 \gg 1$, since in the ET regime $\text{Wi} \gg 1$ and $\text{Re} \ll 1$. The predicted range of $\alpha$ is found in a good accord with the experimental results [6, 20, 21, 23, 40–42], whereas the scaling exponent $\nu$ of the elastic energy spectrum is difficult to verify experimentally. However, numerical simulations at moderate Re and $\text{Wi} \gg 1$ found $\nu \approx 2$ that is different from the value obtained from $\nu = (\alpha - 2) \approx 1.5$ for the experimentally obtained value $\alpha \approx 3.5$, though does not contradict to the prediction (9). Furthermore, the additional power-laws obtained experimentally for torque and pressure power spectra $\beta = 3$ and $\mu \approx 4$, respectively, can be also tested via scaling relations to $\alpha$ derived by scaling arguments [23]. Thus, one gets $\beta = 2(\alpha - 2)$, $\nu = (\alpha - 2)$, and $\mu = 2\nu$. The obtained scaling relations provide reasonable values for $\beta = 3$, $\nu = 1.5 > 1$, and $\mu = 3$, where the last value is less than experimental [23]. The detailed comparison with experiments is presented in Refs. 6, 23.

### 3. Elastic instability and ET of polymer fluids in channel flow with obstacles

#### *3.1. A viscoelastic flow between two widely spaced obstacles*

To understand and characterize a flow between two obstacles hindering a channel flow of highly elastic fluids at $\text{Wi} \gg 1$ and $\text{Re} \ll 1$, we choose a planar quasi-2D channel flow geometry to avoid additional complications due to 3D instability detected experimentally [43]. Besides the control parameters used for all viscoelastic flows, two geometrical parameters: the confinement $a = h/w$ and blockage $b = 2R/w$ ratios, are used. The parameter $b$ controls the relative strength of shear near the wall and extension near a stagnation region, whereas $a$ controls the 2D versus 3D effects. For both small $a \ll 1$ and $b \ll 1$, one expects mainly 2D confined unbounded flow with large extensional strains between two cylinders. Here $h$, $w$, and $R$ are the height, width, and obstacle radius, respectively. In spite of numerous experimental studies, there is no a single one, where an elastic instability and flow above it in a transition regime are properly characterized in flow past either single or array of obstacles hindering a channel flow at $a \ll 1$ and $b \ll 1$. However, in numerical simulations, the first step towards the elastic instability characterization in 2D viscoelastic creeping flow is made in [44].

A flow between two widely spaced obstacles obstructing a channel flow is characterized in [17] by several flow properties: friction factor, pressure and span-wise velocity fluctuations and their power spectra, average stream-wise velocity and wall-normal vorticity, spatial velocity field, and spatial distribution and radially averaged circulation as function of Wi. To conduct these measurements the following techniques are used: pressure drop and fluctuations measurements by various high-resolution pressure sensors, instantaneous flow discharge, 2D particle image velocimetry (PIV), and streak flow imaging. As the result, two continuous linear instabilities at close Wi are characterized: the Hopf bifurcation at $\text{Wi}_{c1}$ due to spontaneous time-reversal symmetry breaking followed by the forward bifurcation at $\text{Wi}_{c2} > \text{Wi}_{c1}$ due to mirror symmetry breaking [17]. The emerging of a small vortex pair close to the upstream obstacle surface at $\text{Wi} < \text{Wi}_{c1}$, the first signature of the flow separation and appearance of the wake, precedes the Hopf bifurcation. The linear Hopf bifurcation leads to spontaneous oscillations of the vortex pair in span-wise direction across the central line. The Hopf bifurcation is characterized by span-wise velocity and pressure fluctuations growth as $(\text{Wi} - \text{Wi}_{c1})^{0.5}$ and Hopf oscillation frequency linear increase with $(\text{Wi} - \text{Wi}_{c1})$ [Figs. 6(a), (b)] in [17]), rms span-wise velocity fluctuations and the average stream-wise velocity (Fig. 5 in [17]), wall-normal vorticity, and friction factor growth as $(\text{Wi} - \text{Wi}_{c2})^{0.5}$ (Fig. 3(b) in [17]), and the linear elongation of the stationary vortex with Wi [42, 17]. The latter feature is unexpected, occurred due to a growing length $l$ at a constant vortex radius and so at a fixed hoop stress resulting in a fixed bulk force $F_h$ [17]. Therefore, since the vortex length $l \sim (\text{Wi} - \text{Wi}_{c2})$ at $\text{Wi} \geq \text{Wi}_{c2}$, so $F_h \sim u^2/lr^2 \text{const}$ due to $u^2 \sim (\text{Wi} - \text{Wi}_{c2})$ [17]. Thus, the striking result of the growing vortex length above the secondary instability is the formation of two mixing layers of parallel shear flow with non-uniform span-wise velocity profiles [42]. Later on, I will use these flows as a paradigm of parallel shear viscoelastic flow dynamics at $\text{Wi} \gg \text{Wi}_c$.

However, the most unexpected result, directly related to geometry of the flow between two widely spaced obstacles, is the next instability occurred above an intermediate transition region characterized by large scatter and slight reduction in $f/f_{\text{lam}}$ (and similar for $\bar{\omega}$) [Fig. 3(b)]. Indeed, the extension of two elongated counter-propagating

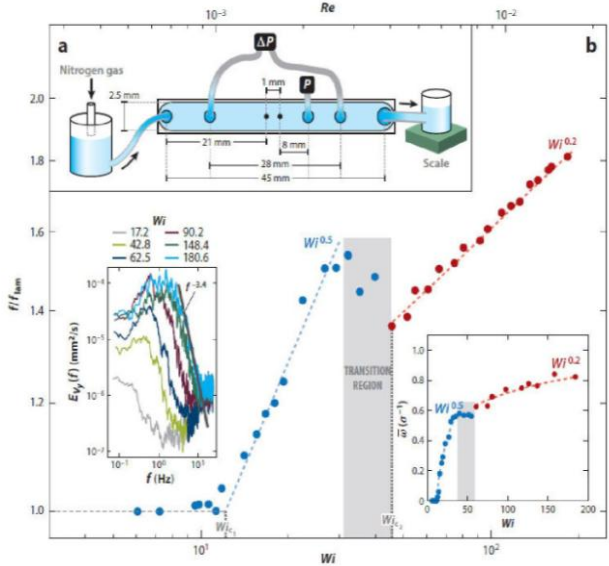

*Fig. 3.* (Color online) (a) Experimental setup of a channel with two obstacles. $\Delta P$ and $P$ are differential and absolute pressure sensors, respectively. (b) Normalized friction factor, $f / f_{\mathrm{lam}}$, versus Wi with dashed lines as fits: $f / f_{\mathrm{lam}} \sim \mathrm{Wi}^{0.5}$ above the Hopf instability, and $f / f_{\mathrm{lam}} \sim \mathrm{Wi}^{0.2}$ in ET with $\mathrm{Wi}_{c1} \approx 12$ and $\mathrm{Wi}_{c3} \approx 59$, corresponding to two transitions. (*Left inset*) The frequency power spectra of the span-wise velocity $E v_y(f)$ for six Wi values. The dashed line are fits giving $\alpha \approx 3.4 \pm 0.1$ in ET. (*Right inset*) Spatially and temporally ($\approx 100$ s) averaged vorticity $\bar{\omega}$ versus Wi with dashed lines as fits: $\bar{\omega} \sim \mathrm{Wi}^{0.5}$ and $\bar{\omega} \sim \mathrm{Wi}^{0.2}$ in two flow regimes. The gray band indicates the intermediate region.

vortices up to filling a full space between two obstacles results in two mixing layers (see Fig. 4 in [17]). In the transition region, velocity fluctuations are enhanced, and the following instability takes place directly to ET accompanied by elastic waves, discovered here for the first time. ET distinguishes oneself by the power-law decay with $\alpha = 3.4 \pm 0.1$ of the span-wise velocity power spectra $E v_y(f)$ at $\mathrm{Wi} > \mathrm{Wi}_{c3} \approx 59$ (Fig. 3(b) lower inset [42]) with lower frequencies pronounced peaks associated with elastic waves in $E v_y(f)$ in lin-log presentation at $\mathrm{Wi} > \mathrm{Wi}_{c3}$ (Fig. 3(b) upper inset [42]). Furthermore, the friction factor, $f / f_{\mathrm{lam}}$ as well as $\bar{\omega}$) grow as $\mathrm{Wi}^{0.2}$ (Fig. 3) due to small fluctuating vortices in ET (see Figs. 4,2SM, 3SM in [42]) quantified via the circulation growing with Wi (see Figs. 5,5SM in [42]). Moreover, $\mathrm{div}\mathbf{V}(x, y)$ in a 2D central plane is found to be close to zero, indicating that the velocity field is 2D and homogeneous (see Fig. 4SM in [42]). Remarkably, since the instability occurs directly to chaotic flow due to finite perturbations observed at $\mathrm{Wi} < \mathrm{Wi}_{c3}$, with $f / f_{\mathrm{lam}}$ and $\bar{\omega}$ characterized by the exponent 0.2 significantly different from 0.5, it indicates that this is not linear, normal mode bifurcation [46]. Finally, left inset in Fig. 4 presents the dependence of $f / f_{\mathrm{lam}}$ on Wi for three

El values. In spite of more than three orders of magnitude difference in El values, reached by varying solvent viscosity, $f / f_{\mathrm{lam}}$ exhibits three dependencies on Wi corresponding to three flow regimes. They are prominent for El = 149 and 2433: transition, ET, and DR, but in particular a complete relaminarization for El = 2433 [47], which is distinctly different from the well-known turbulent drag reduction (TDR) [48] and reported for the first time.

Another important issue investigated in this flow geometry is the role of inertia on the viscoelastic flow stability in a broad range of the control parameters El, Wi, and Re [47]. The corresponding stability diagram of various flow regimes in Wi–Re coordinates reveal non-monotonic dependence on El of the elastic instability as well as transitions to ET and DR regimes (see Fig. 3 in [47]). Then one identifies three distinct regions of El values in the stability diagram. At high El between 14803 and 1070, DR down to a complete flow relaminarization, different from turbulent drag reduction, is discovered at Re < 1 together with the transition and ET regimes (Fig. 4). At $31 \leq \mathrm{El} \leq 304$, in an intermediate region, the same flow regimes are observed but DR does not reach the relaminarization. At El between 9.9 and 1.4, in low elasticity region, two regimes of increasing flow resistance: transition and ET in Wi range between 25 and 1500 and with significant inertial effect at 10 < Re < 300 are found. The third intermediate short regime between the transition and ET is characterized by about a constant value of the friction factor (Figs. 4 and 3 in [47]).

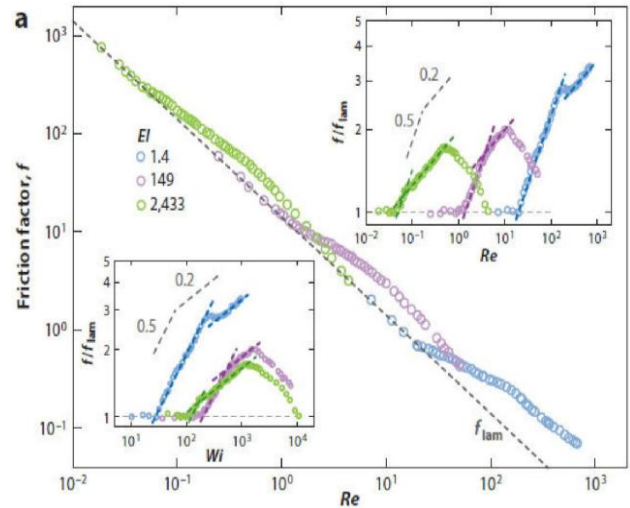

*Fig. 4.* (Color online) Friction factor $f$ plotted against Re for three values of $El$. The dashed line $f_{\mathrm{lam}} \sim 1/\mathrm{Re}$ represents the laminar flow. (*Top inset*) Normalized friction factor $f / f_{\mathrm{lam}}$, versus Re. (*Bottom inset*). The same data presented as $f / f_{\mathrm{lam}}$, versus Wi fitted by dashed lines in transition $f / f_{\mathrm{lam}} \sim \mathrm{Wi}^{0.5}$ and ET $f / f_{\mathrm{lam}} \sim \mathrm{Wi}^{0.2}$ flow regimes. The drag reduction at Re $\approx 0.5$, Wi $\approx 1{,}216$, and El = 2,433 proceeds down to complete relaminarization.

*3.2. Elastic waves in elastic turbulence*

Using the similarity of MHD and viscoelastic equations based on the Oldroyd-B polymer model and describing isotropic, homogeneous, incompressible and random flow (1)–(4), presented in uniaxial form in ET (5)–(8), the elastic waves are predicted [38] by analogy with the Alfven waves in MHD [49, 39]. Indeed, from Eqs. (5)–(8) and by considering small perturbations $U(x,z,t) = U_x(z) + u(z,t)$ and $B(z,t) = B_x x + b(z,t)$ in a span-wise components of $U(x,z,t)$ and $\sigma_{xx} = B_x B_x$, one gets the linear equations of transverse stream-wise propagating elastic waves in 2D ($x$–$z$) plane at Wi >> 1 and Re << 1,

$$\partial u_x / \partial t = (1-\beta) B_x \partial b_x / \partial z, \quad (10)$$

$$\partial b_x / \partial t = B_x \partial u_x / \partial z, \quad (11)$$

where $U_x(z)$ is stationary stream-wise shear velocity. Then by analogy with the Alfven waves, one gets the elastic wave linear dispersion relation as

$$c_{el} = \omega / k = (\sigma_{xx} / (1-\beta)\rho)^{1/2} = B_x / ((1-\beta)\rho)^{1/2} \quad (12)$$

or generally in 3D

$$\omega = (\mathbf{k}\hat{\mathbf{n}})[tr(\sigma_{ij})/\rho]^{1/2} = (\mathbf{k}\hat{\mathbf{n}}) c_{el}, \quad (13)$$

where $c_{el}$ is the elastic wave speed providing unique information about the elastic stress $\omega$ and $\mathbf{k}$ are the rotation frequency and wave vector of the elastic waves, respectively, and $\hat{\mathbf{n}}$ is the director that defines a major stretching of $\sigma_{ij}$ and elastic wave propagation direction [38]. A simple physical explanation of both Alfven and elastic waves follows from an analogy between the response of either magnetic or elastic tension on transverse perturbations and an elastic string when plucked. Thus, both the Alfven and elastic waves are transverse to the propagation direction, unlike longitudinal sound waves in plasma, gas, and fluid media, and reveal linear dispersion [16]. Moreover, $c_{el}$ provides unique information about the elastic stress (13), whereas the wave amplitude is proportional to the transversal perturbations of elastic stress (11), otherwise unavailable experimentally.

As mentioned above, the first evidence of elastic waves is obtained from well-pronounced low frequency peaks in the span-wise velocity power spectra at Wi > $\text{Wi}_{c3}$ to ET, where the peaks represent elastic wave [Fig. 3(b)], since its frequency grows nonlinearly with Wi [Fig. 5(b) inset]. The appearance of low frequency peaks in the span-wise velocity power spectra above the first linear instability at Wi > $\text{Wi}_{c1}$ are the Hopf oscillations and their frequency linearly scales with Wi – $\text{Wi}_{c1}$ (Fig. 6(a) and 6(b) in [45]). The main experimental proof of elastic wave finding is the dependence of $c_{el}$ on Wi. Since a technique to measure elastic stress in a random flow is currently absent, it is the only way to get the dependence of $c_{el}$ on Wi and relate it to the elastic stress dependence via a model. By exploring temporal cross-correlation function $C_v(\Delta x, \tau)$ of stream-wise velocity $u(x,t)$ calculated for two spatially separated points on a central line $\Delta x$, one gets a time shift $\tau_p$ of $C_v(\Delta x, \tau)$ for each $\Delta x$. Then by varying $\Delta x$, a family of $C_v(\Delta x, \tau)$ is obtained [Fig. 5(a)] [50], and a linear dependence of $\Delta x$ on $\tau_p$ provides the elastic wave speed $c_{el} = \Delta x / \tau_p$ for a given Wi (Fig. 5(a) inset for Wi = 148.4). The dependence of $c_{el}$ on Wi is shown in Fig. 5(b) together with the fit that gives the scaling relation $c_{el} = A(\text{Wi} - \text{Wi}_{c3})^{\xi}$, where $\xi = 0.73 \pm 0.12$, $A = 5 \cdot 10^{-3}$ m/s, and $\text{Wi}_{c3} = 59.7 \pm 1.8$, close to that found from $f/f_{lam}$ and $\bar{\omega}$ dependencies on Wi [50]. This scaling expression is confirmed in several parallel shear flow geometries: in (i) a flow past an obstacle [51], (ii) a straight channel flow with [52] and without [53] strong prearranged perturbations at the inlet and stream-wise propagating waves; and (iii) in span-wise propagating elastic waves in a weakly perturbed channel flow [54]. It is important to emphasize that the expression (13) is based on the Oldroyd-B linear polymer elasticity model [38]. Then using the expression for the first normal stress difference, $N_1$, one gets $c_{el} = [\text{tr}(\sigma_{ij})/\rho]^{1/2} = (N_1/\rho)^{1/2} = (2\text{Wi}^2\eta/\lambda\rho)^{1/2}$ [4] that gives $c_{el} = (2\eta/\lambda\rho)^{1/2}\text{Wi}$. It differs from the scaling expression following from the experiment presented above in both the scaling exponent and the coefficient value. This discrepancy results from the Oldroyd-B linear model used by theory [38].

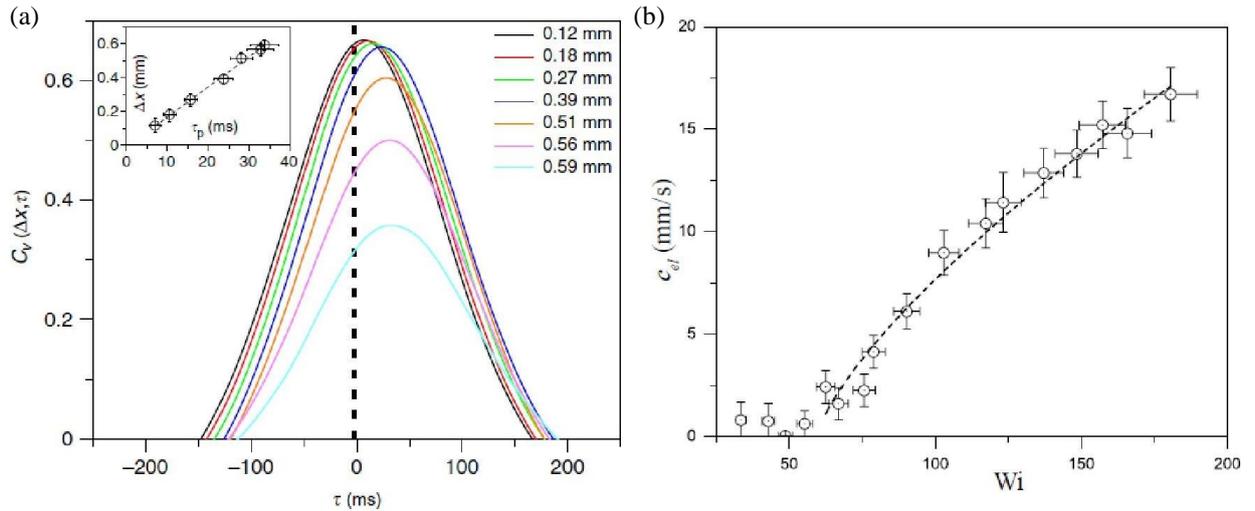

*Fig. 5* (Color online) (a) Cross-correlation functions of the stream-wise velocity $C_v(\Delta x, \tau)$ versus lag time $\tau$ for different values of $\Delta x$, obtained at $y/R = 0.18$ and for Wi = 148.4. Inset: $\Delta x$ versus $\tau_p$ at Wi =148.4, and a slope of linear fit (shown by dashed line) gives $c_{el}$ with error bars for $\Delta x$ determined by the spatial resolution and for $\tau_p$ based on the width of Gaussian fit of $C_v(\Delta x, \tau)$. (b) Dependence of $c_{el}$ on Wi, where the dashed line is a fit $c_{el} = A(\text{Wi} - \text{Wi}_{c3})^{0.73 \pm 0.12}$ that yields $\text{Wi}_{c3} = 59.7 \pm 1.8$.

Remarkably, the elastic waves are not observed in flow geometries with curvilinear streamlines in either above the instability in transition or ET regimes. Moreover, our early attempts to excite the elastic waves in either laminar elongation flow, which contrary to shear flow distinguishes by strong polymer stretching, or flows with curved streamlines in ET at Wi >> 1 and Re << 1 were unsuccessful. The only suggested explanation is that transverse periodic oscillations used in the experiments or transverse nature of the elastic waves with curved streamlines result in additional hoop stress generating a bulk force directed towards a curvature and suppressed them.

*3.3. Mechanism of relaminarization of elastic turbulence*

Relaminarization of ET, firstly observed in the flow between two obstacles obstructing a channel flow (Fig. 4) and later in a flow past an obstacle (Fig. 2 in [51]) at El >> 1, is recently detected in various viscoelastic straight channel flows and never observed in viscoelastic flows with curved streamlines. Moreover, both elastic waves and relaminarization are discovered only in channel flows of dilute polymer solutions, and the mechanism of the DR and relaminarization is generic and universal for all parallel shear flows of viscoelastic fluids in ET.

A flow past an obstacle is a paradigmatic problem of fluid mechanics for both Newtonian and viscoelastic fluids widely investigated in the past both experimentally and numerically. Studies of highly elastic fluids at Wi >> 1 and Re << 1 in such a flow geometry are rather limited. The key role of elastic waves in relaminarization at Wi >> 1 and Re << 1 are the subject of a recent paper, where this problem is studied in a flow past an obstacle [51], similar to the flow between two obstacles discussed above. In contrast to the latter, the flow past an obstacle hindering a channel flow differs by an absence of mirror symmetry in a stream-wise direction in the downstream wake. Since in this study we use a wider range of El up to the highest El = 28251, DR is found for all four values of El, whereas relaminarization is observed only for the two highest El values from them (Fig. 2 in [51]). In the paper, we discuss and resolve two problems: (i) What is the enhancement mechanism of the friction factor, $f/f_{lam}$, in ET above the secondary non-modal bifurcation? (ii) What is the reason for the decay of elastic wave intensity in ET? Regarding (i), we demonstrate experimentally a similar growth and reduction with Wi at four values of El for the following flow properties: $f/f_{lam}$, $I/v_{rms}^2$, $P_{rms}$, $v_{rms}/U$, and surface area and length of the wake. Moreover, these flow properties change the tendency from growth to decay at about the same Wi value corresponding to ET-to-DR transition, $\text{Wi}_c^{DR}$. Here $I/v_{rms}^2$, $P_{rms}$, and $v_{rms}/U$ are the normalized intensity of elastic waves, rms pressure fluctuations, and the rms span-wise velocity normalized by the average stream-wise velocity, respectively. The first two properties as function of Wi are presented in Figs. 6(a)–(c), and the dependence of $f/f_{lam}$ on $I/v_{rms}^2$ in both ET and DR are shown in Fig. 7(a), (b). In addition, wall-normal vorticity in streak flow images agree well with the observed correlation in both regimes (see Fig. 1 and the same image in S1 at El = 2851 in [51]).

The main observation for understanding the mechanism of the ET attenuation leading to DR and relaminarization is an excellent correlation between $f/f_{lam}$ and $I/v_{rms}^2$ (Fig. 7). Indeed, the larger (smaller) $f/f_{lam}$, the greater (smaller) $I/v_{rms}^2$, and $I/v_{rms}^2$ tends to zero at relaminarization, quantified in their dependence. Thus, we suggest that synchronous interaction of the elastic waves with wall-normal vorticity fluctuations leads to their amplification in ET and subsequent suppression in DR. This mechanism of the resonant interaction results in an effective energy pumping from the elastic waves to wall-normal vortex fluctuations. The physical mechanism of the interaction of the elastic waves with fluctuating vortices is analogous to the Landau wave damping [52], occurred due to the resonant interaction of electromagnetic waves with fast electrons in plasma, when the electron velocity close to the wave phase speed. Similarly, acoustic damping occurs in sound-gas bubble interaction [53] resulting in strong wave attenuation. Second, the main reason for drastic change from $I/v_{rms}^2$ growth in ET to its decrease in DR is related to the strong increase in dissipation of elastic waves caused by its frequency $\nu_{el} = \omega/2\pi$ growth. Estimates of the range of the elastic wave frequencies with low attenuation limited by two dissipation mechanisms [54] at El = 11232 give $\nu_{el} \approx 8$ Hz, above which the attenuation strongly increases in reasonable agreement with the experiment (see Fig. S5(a) in [51]). Thus, the appearance of large attenuation of elastic waves at higher frequencies is the reason for the emergence of DR and relaminarization.

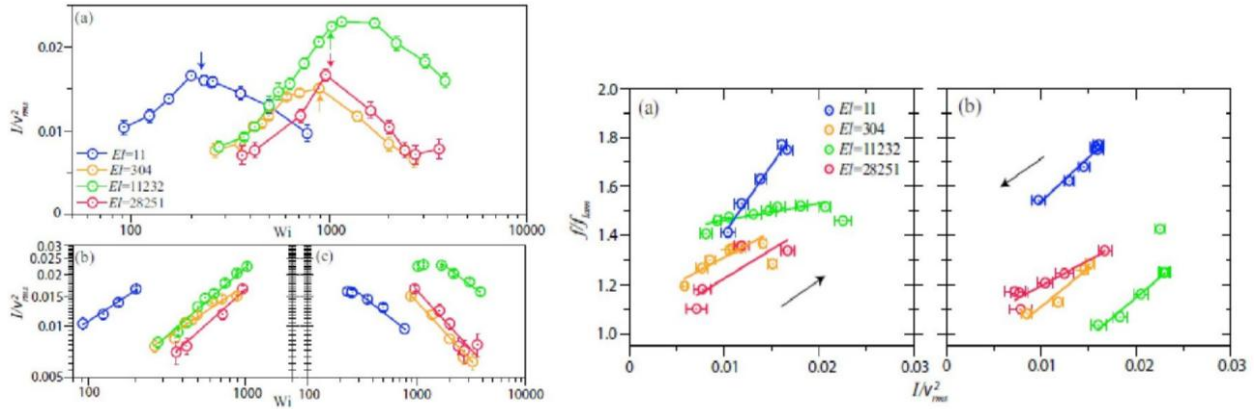

*Fig. 6.* (Color online) ($I / v_{rms}^2$) versus Wi in (a) linear and (b), (c) log–log scales exhibiting power-law in ET and DR regimes for various El. Arrows indicate the transition values of Wi from the ET to DR.

*Fig. 7.* (Color online) Correlation between the friction factor and normalized elastic wave intensity in (a) ET and (b) DR, respectively. Arrows in (a) and (b) indicate direction of increasing Wi. Solid lines are linear fit.

## 4. Non-modal instability, coherent structures, and elastic turbulence in viscoelastic channel flows with strong and weak perturbations at inlet

### 4.1. Non-modal bifurcation in parallel shear flows and coherent states

As pointed out in Introduction, the instability mechanism of viscoelastic flows with curved streamlines becomes ineffective in parallel shear flows, and their linear stability is proved. The latter does not imply their global stability. One of such mechanisms is revealed in Newtonian parallel shear flows, which are known to be linearly stable, though all of them exhibit transition to turbulence flow at finite Re [46]. A linear stability of these flows is described by linearized Navier–Stokes equations called Orr–Sommerfeld equation, which involves non-self-adjoint (non-Hermitian) operator [55, 56]. As a result, a linear stability analysis is made only by using a self-adjoint approximation. However, linear modes are stable: after initial exponential growth, they decay exponentially. Therewith, non-normal modes of the non-self-adjoint operator may become unstable, grow algebraically up to large amplitudes exceeding the linear modes, though they are non-orthogonal to each other resulting in non-unique numerous instability modes. Thus, the non-normal mode instability involves a combination of several such modes [55, 56]. It means that finite-size random perturbations expected to be experimentally relevant. Moreover, unstable linearly growing non-normal modes may emerge from general random perturbations but a selection of the most stable non-normal modes would be sensitive to a perturbation spectrum [57]. An algebraic growth in time of non-normal modes provides another pathway to amplify small finite-size perturbations by many orders of magnitude, sufficient to excite the nonlinear mechanism stabilizing the non-normal modes [55, 56]. Thus, a linearly stable flow becomes unstable due to finite-size perturbations. Remarkably, contrary to a normal mode bifurcation, a non-modal instability onset is not uniquely defined and depends on the amplitude of localized perturbations. An experiment verified the $Re^{-1}$ dependence of the critical normalized amplitude $\Phi_c(Re)$ of a single localized perturbation of a Newtonian pipe flow, promoting a transition to turbulence [58]. This result means that for each $Re < Re_c(\Phi)$ an initial condition with a given $\Phi$ eventually settles into laminar flow, and at $Re > Re_c(\Phi)$, a new flow state develops due to a non-modal instability.

Thus, a flow structure above the non-modal instability depends on the details of the perturbations and is highly sensitive to their amplitude and spectrum [57]. Indeed, in channel, pipe, and plane Couette flows, a span-wise modulated flow is selected from a multitude of the most amplified structures [59]. Indeed, the transition to turbulence is characterized by appearance of span-wise modulated exact coherent structures (ECSs) attributed to non-normal weakly unstable modes. ECSs are self-organized into a cycling self-sustained process (SSP). This bypass route to turbulence is discovered numerically and verified experimentally in Newtonian parallel shear flows during the last two decades [55, 59]. Thus, due to the non-normal instability, weakly unstable ECSs, stream-wise vortices and streaks, are found to be a persistent feature in both numerical simulations [55, 59] and experiments [60].

Similarly to the non-modal analysis of linearly stable Newtonian shear flows at Re >> 1, the non-modal approach to instability of viscoelastic channel and pipe inertia-less flows at Wi >> 1 (and El >> 1) is introduced in [61–63]. Since the theory is based on the linearized Navier–Stokes and elastic stress dynamics equations for the non-normal modes, it describes only transient stages of the transition and is not able to predict ECSs prevailed in a regime of the nonlinear saturation. Several studies have experimentally revealed the elastic instability that occurs in inertia-less straight shear flows due to strong prearranged perturbations, generated by either jet [11] or obstacle array [12–14] at the inlet, contrary to non-modal instabilities studied in Newtonian parallel shear flows. The elastic transition in pipe [11] and straight square micro-channel [12–14] flows of dilute polymer solutions at Wi >> 1 and Re << 1 directly to a chaotic flow, similar to Newtonian shear flows, are reported. Above the elastic instability, which was not characterized properly, both flows are distinguished by continuous velocity power spectra with power-law decays at higher frequencies with slope exponents of about, $\alpha \approx -1.5$ [11] and from $\alpha \approx -1.7$ down to $-2.7$ [12, 13]. However, in both cases, ET is not reached in spite of the claims.

### 4.2. Discovery of non-normal mode bifurcation, coherent states, and elastic waves in a channel flow with strong perturbations

A recent experiment from our lab studied the elastic instability and characterized the flow in a quasi-2D straight channel [64, 65]. Similarly to [12–14], the flow was strongly perturbed by an array of obstacles across the full channel width at the inlet (Fig. 8). Three flow regimes are observed above the instability onset: transition, ET, and DR, where the bifurcation was found to be continuous and non-hysteretic (see Fig. 2 E, F in [64]). Moreover, it is observed that the instability is a non-normal mode bifurcation, as determined by examining three different features. First, the Wi dependencies of $C_f / C_f^{\text{lam}}$, $u_{rms}/U$, and $P_{rms}$ have slope exponents about 0.125, 0.95, and 0.2, significantly different from the indicative value 0.5 for the normal-mode bifurcation (Fig. 9). Second, just above the instability onset, a continuous velocity power spectrum with an algebraic decay at higher frequencies, in addition to high-energy peaks in the span-wise velocity spectrum at lower frequencies are found [64]. Thus, an infinite number of modes are excited above the instability threshold, contrary to the single most unstable mode in the case of the normal mode instability [46] (Fig. 10). Third, the spectral peaks at low frequency [Fig. 10 (right)] implies the existence of elastic waves on top of a chaotic flow in the transition, ET, and DR regimes. The dependence of the elastic wave speed on Wi, similar to that shown in Fig. 5(b), verifies the elastic waves. These features confirm that the instability is the non-normal mode bifurcation excited by finite-size perturbations. Furthermore, in all three regimes, weakly unstable coherent structures (CSs), namely stream-wise vortices (rolls) and streaks, similar to ECSs in the Newtonian case, are discovered via PIV [Fig. 11 (left)]. Moreover, CSs are accompanied by elastic waves emerging above the onset and playing the key role in supporting CSs. As a result, CSs are self-organized into a cycling SSP synchronized by the elastic waves that is proved in Fig. 11 (right) by presenting two consecutive cycles in ET. It is also verified that due to Wi increase, the CSs and their dynamics depend on the intensity of the elastic waves detected in three flow regimes. It clearly reveals in appearing only in ET a secondary elastic instability of the interface counter-propagating streaks (in a frame moving with $U(z)$) due to higher elastic wave energy, resulting in temporary dynamics similar to Kelvin–Helmholtz instability, though due to strikingly different elastically driven mechanism, suggested in [65]. Thus, the observed similarity in the flow dynamics in Newtonian and viscoelastic parallel shear flows suggests the universality in stochastically steady state of ECSs, self-organized into cycling SSP in random states that occurred via non-normal mode instability, otherwise linearly stable flows. The only difference between Newtonian and viscoelastic parallel shear flows is the observation of CSs and elastic waves only in a limited part of the channel length up to $l/h \approx 200$, probably, due to significant elastic wave attenuation downstream from perturbation source.

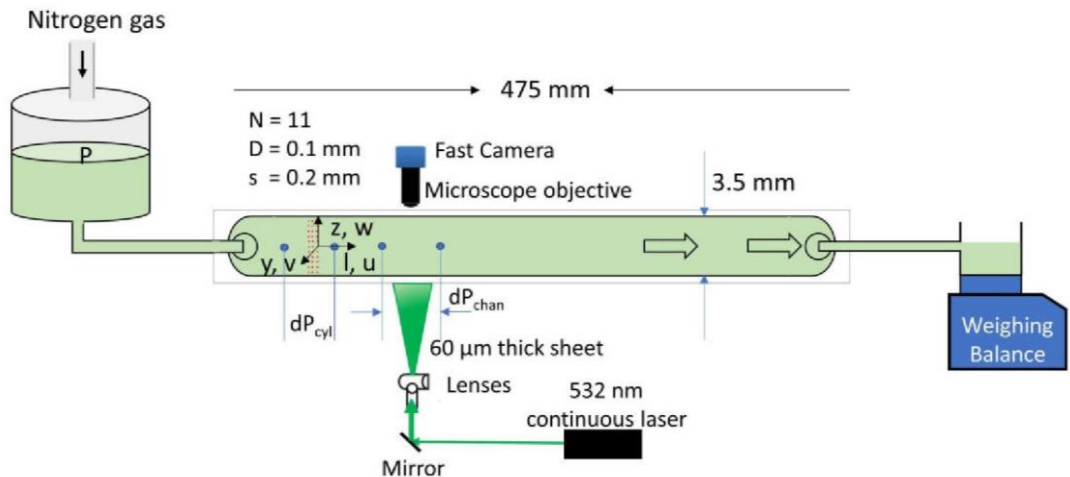

*Fig. 8.* (Color online) Experimental setup. Straight quasi-2D channel of $l \times w \times h$ = 475×3.5×0.5 mm has an array of 3×11 cylindrical obstacles of 100 μm with 200 μm apart to generate strong perturbations at the inlet. Two pressure sensors measure the pressure drop, $\Delta P_{cyl}$ and $\Delta P_{chan}$, across the obstacles array and channel, respectively, and the absolute pressure sensor measures $P_{rms}$. The Nitrogen gas, pressurized up to 100 psi, drives the polymer solution. Velocity field is measured via PIV. PC-interfaced balance weights fluid at the outlet as a function of time.

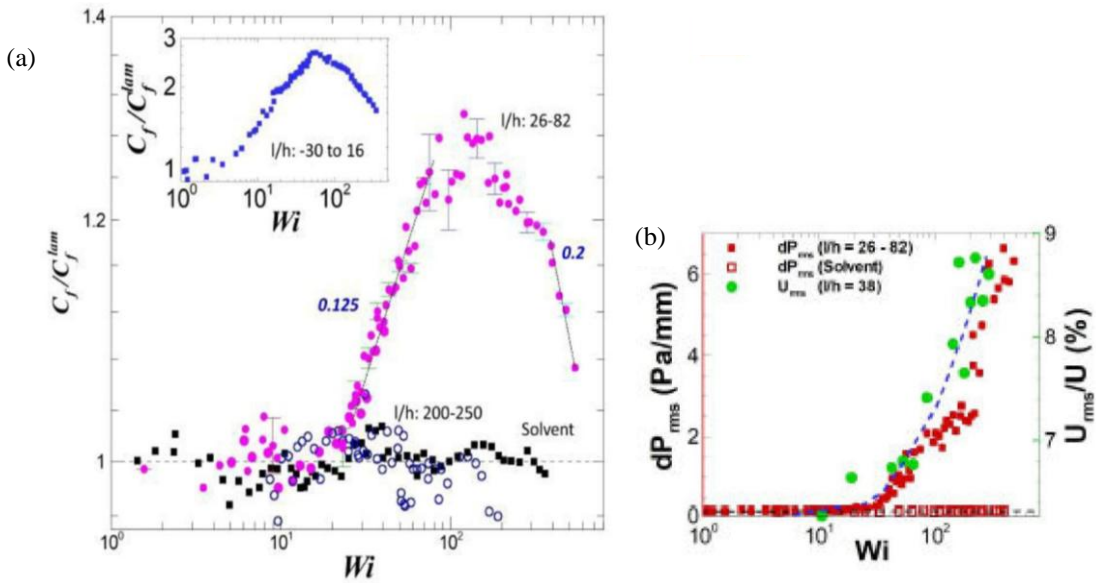

*Fig. 9.* (Color online) (a) Friction factor $C_f / C_f^{\text{lam}}$ versus Wi for three distances from the third cylinder row (at $l/h = 0$) $l/h$: (i) –30 to 16 (inset), (ii) 26 to 82 (filled circle), (iii) 200 to 250 (open circle), and Newtonian solvent at $l/h = 26$ to 82 (filled square). Inset shows $C_f / C_f^{\text{lam}}$ versus *Wi* across three rows of cylinders. (b) The dependence of $dP_{rms}$ at $l/h = 26$ to 82 and $u_{rms}/U\,(\%)$ on Wi at $l/h = 38.6$. One notices growth for all variables above the transition. (Wi $= U\lambda/w$, $w = 3.5$, $h = 0.5$, and $l = 475$ mm are the channel width, height, length.)

Further, one finds only the normalized stream-wise velocity fluctuations up to 4% downstream until the outlet [64].

*4.3. Non-normal mode bifurcation, coherent states, and elastic waves in a channel flow with weak perturbations*

Next two experiments are designed to elucidate the following questions related to the results obtained in the first experiment in a straight channel flow of viscoelastic fluid [64].

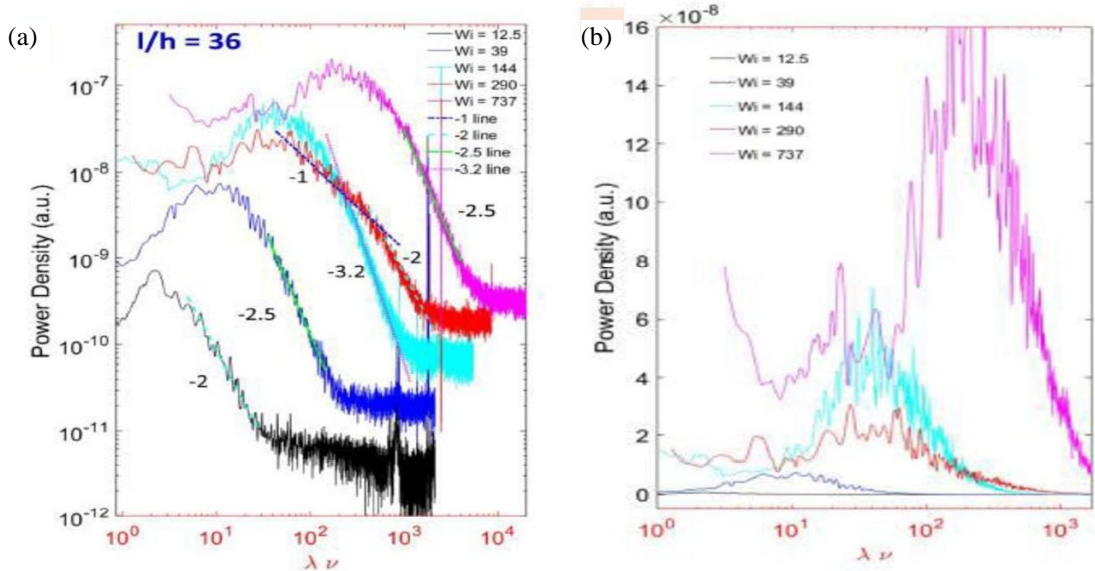

*Fig. 10.* (Color online) (a) Span-wise velocity power spectra for various Wi at $l/h = 38.6$. At larger frequencies, all spectra show the power-law decay with exponents depending on the flow regime: at $\alpha = 2$ up to 2.5 transition, at $\alpha = 3.2$ ET, back to 2 DR, then rise back to 2.5 above DR. (b) Span-wise velocity power spectra in lin-log scale to compare the maximum values of energy peaks at several Wi in different flow regimes.

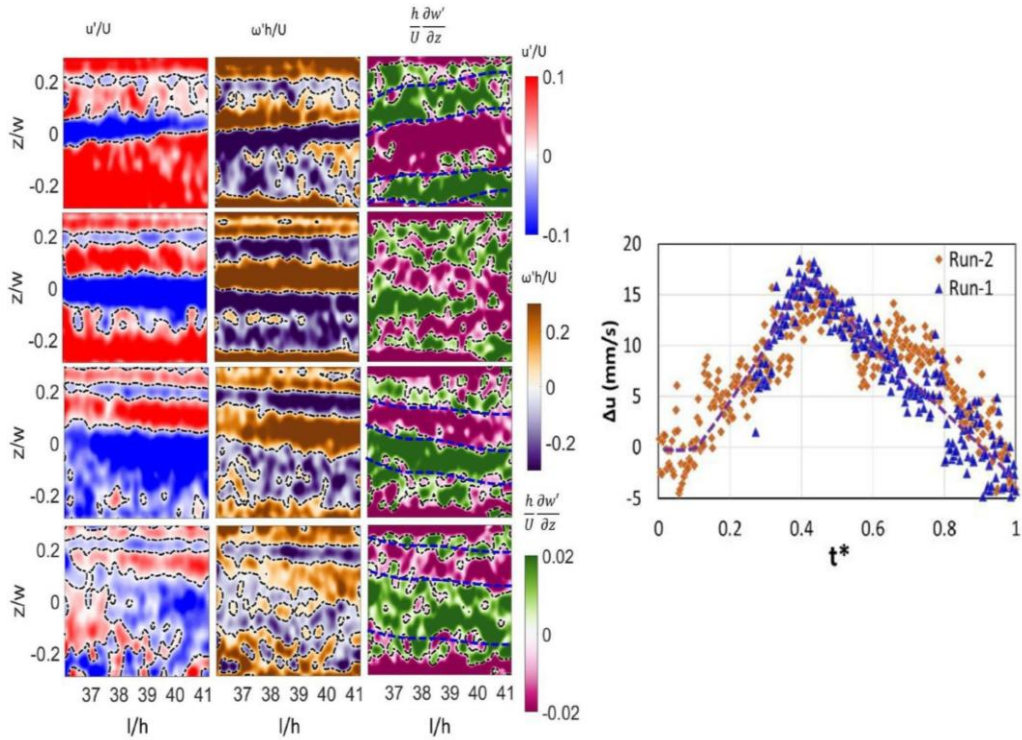

*Fig. 11.* (Color online) (Left) Cycling coherent structures in ET at Wi = 185 and $l/h$ = 36–41. The normalized time $t^* = tf_{el}$, where $f_{el}$ is the elastic wave frequency. Fluctuating stream-wise velocity $(u'/U)$, vertical vorticity $(\omega'h/U)$, and span-wise gradient of span-wise velocity $(h/U \partial w'/\partial z)$ are shown. Flow structures at four values $t^*$ in a cycle are shown in four rows for $t^*$ = 0.15, 0.29, 0.56, and 0.67 (from the top). At $t^*$ =0.15, CS of stream-wise streaks of high and low speed are shown as positive and negative $u'/U$ separated by dotted black lines of $u'/U$ exhibiting span-wise fluctuations. In $(\omega'h/U)$ plot, strips of wall-normal vorticity at the center of streaks are identified together with its random pattern near the edge of strips. The contour of $\partial w'/\partial z$ presents stream-wise rolls, another CS, together with a random structure (see for explanation Fig. S1 and Suppl. Notes (SI) in [64]). At $t^*$ = 0.29, CSs of $u'/U$ and $\omega'h/U$ look even more pronounced, whereas $\partial w'/\partial z$ field is significantly perturbed. At $t^*$ = 0.56, streaks and wall-normal vorticity CS are strongly perturbed, but the rolls is more pronounced in $\partial w'/\partial z$ marked by dotted blue lines and at this time both CSs co-exist. Finally, at $t^*$ = 0.67, all fields are chaotic, and later, a new cycle starts. (Right) Temporal evolution of $\Delta u = u_4 - u_2$, where $\Delta u$ is the velocity difference at two specific locations across the interface (see Figs. S10, SI in [64]), is shown for two consecutive cycles in ET at Wi = 185. Their periods and the maximum values of $\Delta u$ are close, and the shape variations during the cycles are surprisingly similar.

(i) Are the strong perturbations at the inlet necessary to get an elastic instability in such flows? (ii) Would the flow exhibit the same characteristics and properties of the non-normal mode bifurcation as reported in [64]? (iii) If yes, would the flow structure be similar to the strongly perturbed flow, namely, three flow regimes with similar CSs, accompanied by elastic waves? (iv) Would there exist the correlation between elastic wave intensity and flow properties as a function of Wi in three flow regimes?

### 4.3.1. Channel flow with smooth inlet and weak perturbations by a single cavity

The first experiment [66] is conducted in the setup shown in Fig. 12. Due to carefully smoothed and tapered over a distance of roughly $200H$ channel inlet to eliminate any unwanted flow disturbances, an elastic instability at very high Wi = $\rho UH/\eta$ and far away downstream of the cavity is observed. It is initiated by finite-size small perturbations caused by a flow inside the cavity located at a middle of the channel. Here $H$ is the channel height. Indeed, a small but deep cavity ($D \cdot l$ = 0.5×5 mm) drilled at the top plate, initially served as a port for pressure measurements, generates the elastic instability at significantly higher Wi (Fig. 12) than by strong perturbations. In spite of much weaker finite-size fluctuations in the transition regime [64], we were able to measure normalized stream-wise velocity $u_{rms}/U$ and pressure $P_{rms}/P_{rms}^{lam}$ fluctuations. They exhibit the scaling dependences on Wi above the onset of the forward non-modal bifurcation, similar to those found in [64]. The velocity and pressure power spectra also reveal similar scaling exponents in three flow regimes: transition, ET, and DR. Furthermore, we observe only streaks, which are so weaker than in [64] that it does not allow to detect their cycling frequency. Thus, these flow properties downstream of the cavity at Wi > $Wi_c$ in the transition regime confirm that the instability is non-modal, similar to [64]. Both ET and DR also observed only downstream of the cavity, whereas upstream of the cavity the flow remains laminar. Moreover, the streaks are detected only in a part of the channel downstream of the cavity at $40H < S < 210H$, similar to [64], with $u_{rms}/U$ up to 8%, comparable with that found in [64]. Further on, velocity fluctuations reduce to less than 2.5% and streaks are not found at all for $S > 210H$ [66].

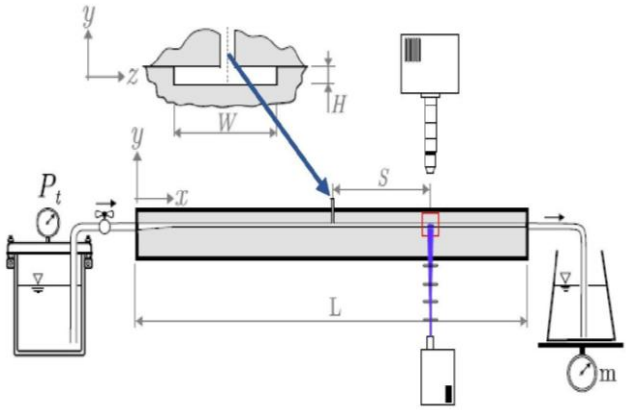

Fig. 12. (Color online) Experimental setup with a cross-section view showing the cavity of diameter $D = H = 0.5$ mm. Quasi-2D straight channel $L \times W \times H = 750 \times 3.5 \times 0.5$ mm, $x$-$z$ mid-plane is located at $y = 0$. All techniques used including PIV arrangement shown in schematics are similar to those described in Fig. 8.

However, the most surprising observation is the elastic wave properties, which are found to be different in several aspects from those detected in [18, 50, 64]. The elastic waves are excited at the instability onset and persists further in ET and DR with the intensity growing in ET and decaying in DR [Fig. 13 (B)]. Due to the higher span-wise against stream-wise gradients of stream-wise velocities fluctuations (see Fig. 9 in [66]), the elastic waves propagate in span-wise direction [Fig. 13 (A) (a), (b)]. As the result of small velocity fluctuations and, in addition, band-pass filtering around the spectral peaks to remove background noise, the space-time plot of the stream-wise velocity spatial structure is plotted in Fig. 13 (A) (a) for Wi = 381, 584, and 808 in transition, ET, and DR flow regimes, respectively, at $S = 170H$. In Fig. 13 (A) (b), the 3D transverse wave structure is shown by the data with phase averaged velocity fluctuations propagating in span-wise direction at Wi = 407, in a contrast with other flow geometries with stream-wise propagating waves [18, 50, 64]. The wave structure in Fig. 13(A) (a), (b) allows to get frequency $f$ and wavelength $l$ of the elastic waves directly from periodicity in $t$ and $z$ coordinates, respectively, and the wave speed $v$ from the angle between the wave crests and $t$ direction as a function of Wi [Fig. 13(C) (a)–(d)]. As seen from Fig. 13(C) (c), $l$ is independent of Wi but with significant fluctuations in $l$ in approximate range of $\epsilon$ [0.5, 0.8] mm. Moreover, from Fig. 13(A) (b), one gets about $2l$ per a half width, or $4l/W \approx 1$, i.e., it seems that the channel width determines the wavelength of the span-wise propagating elastic waves. Therewith, the velocity and frequency of the elastic waves exhibit the same scaling with $Wi - Wi_c$, and the wave velocity shows linear dependence on $fl$ verifying the predicted linear dispersion for the first time [38]. Finally, the fit of the $v$ dependence on $Wi - Wi_c$ provides $v = A(Wi - Wi_c)^\gamma$ with $\gamma = 0.73 \pm 0.05$ and $A = 3 \cdot 10^{-3}$ mm/s. Remarkably, there is an excellent agreement in the value of $\gamma$ obtained in [50] but the A value is almost three orders of magnitude smaller than found in [50, 64]. Thus, due to small rms velocity fluctuations and extremely small wave frequencies, the wave attenuation is orders of magnitude smaller [Fig. 13 (A) (b)] than in [64]. It explains the observation of the elastic waves up to $S = 170H$. However, it is unclear, why $A$ is found so different for span- and stream-wise propagating elastic waves.

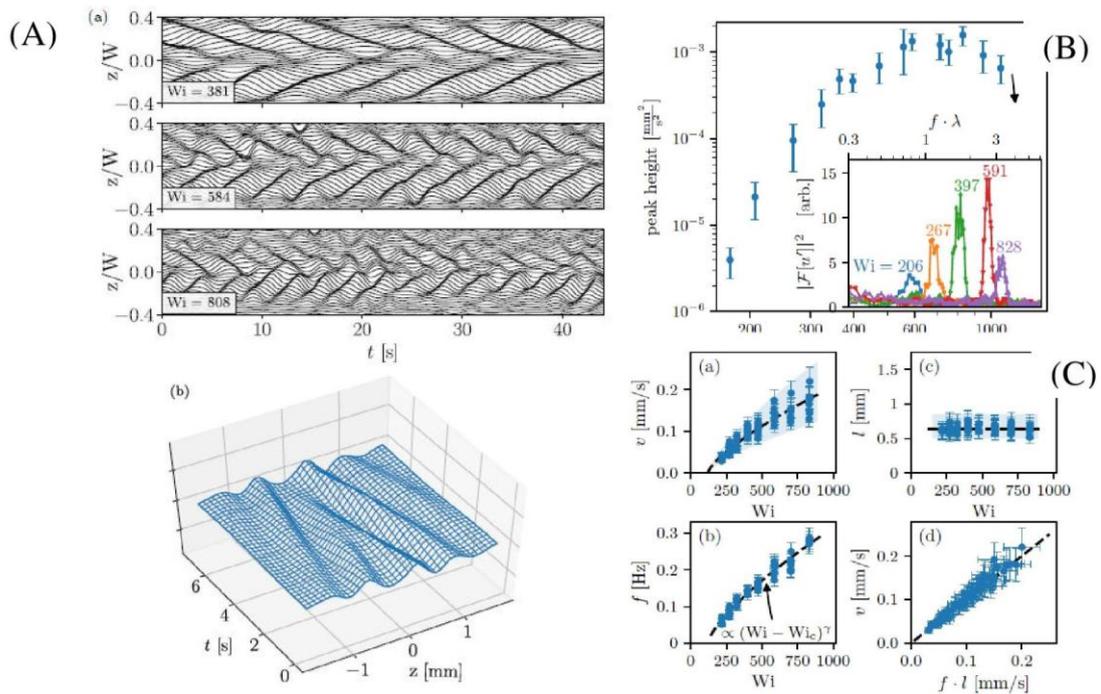

Fig. 13. (Color online) (A) (a) Space-time plots at $-0.4 < z/W < 0.4$ of stream-wise velocity fluctuations, $u'_x(z,t)$, for three Wi values. (b) 3D image of span-wise elastic wave at Wi = 407. Stream-wise velocity is phase averaged. (B) Intensity of the elastic wave versus Wi, in log-log scales. Inset: stream-wise velocity power spectra at $S = 170H$ and $z = \pm H/6$ for five Wi values in lin-log scale. The spectra and frequencies are normalized by the noise level and $\lambda$, respectively. (C) (a), (b), (c), and (d) show the wave velocity $v$, frequency $f$, wavelength $l$, dependencies on Wi, and $v$ versus $f \cdot l$, respectively. Dashed lines are the best fits.

### 4.3.2. Channel flow with non-smooth inlet generating weak perturbations

We conducted another experiment in the setup shown in Fig. 14(a) [67]. The relevant sources of finite-size disturbances are the non-smoothed inlet and six 0.5 mm holes at the top plate for the pressure drops and fluctuations measurements. Otherwise, the channel dimensions, design, gas driven flow, measuring techniques, and polymer fluid properties are similar to those used in [64, 65]. Fig. 14(b) presents high-resolution measurements of the friction factor, $f/f_{lam}$, and Fig. 14(c) shows the normalized pressure and stream-wise velocity fluctuations versus Wi in three flow regimes characterized by the scaling exponents, close to those found in a channel flow with strong perturbations at the inlet [64] (see Fig. 9). Further on, Fig. 15(a) presented stream-wise velocity power spectra versus $\lambda f$ far downstream of the inlet disclose their decay exponents from –1.2 to –2.8 in transition, down to $-3.0 \pm 0.2$ in ET, and back to –2.8 in DR. Thus, the scaling dependence of $f/f_{lam}$, normalized stream-wise velocity and pressure fluctuations on Wi, and stream-wise velocity spectra in the transition regime provide a strong evidence of the non-modal nature of the elastic instability [61–63]. Furthermore, in the inset in Fig. 15(b), an elastic wave peak in span-wise velocity power spectrum in lin-log scales at Wi = 2034 is shown. The values of the peak frequencies, $f_{el}$, are used to obtain the dependence of $c_{el}$ on Wi presented in Fig. 15(b). The scaling of $c_{el}$ with $(Wi - Wi_c)^\xi$ provided by the fit gives $A = (0.45 \pm 0.05) \cdot 10^{-3}$ m/s and $\xi = 0.72 \pm 0.02$ in an excellent agreement with early results [50, 64] and in accord with the span-wise propagating elastic waves in the ξ value [66]. Finally, we observe streaks, the only CS, similar to [64, 66], in three flow regimes. To prove experimentally the periodicity of a cycled SSP with the elastic wave period we utilize the approach developed earlier in [64] by examining a temporal evolution of the velocity fluctuation difference at two specific points across the interface $\Delta u' = u'_2 - u'_1$, as shown in Fig. 15(c), similar to Fig. 11 (right). During each cycle, $\Delta u'$ displays a non-monotonic temporal variation with the same peak value $\Delta u'_{max}$ and cycle period for four cycles, in spite of highly fluctuating flows [Fig. 15(c)].

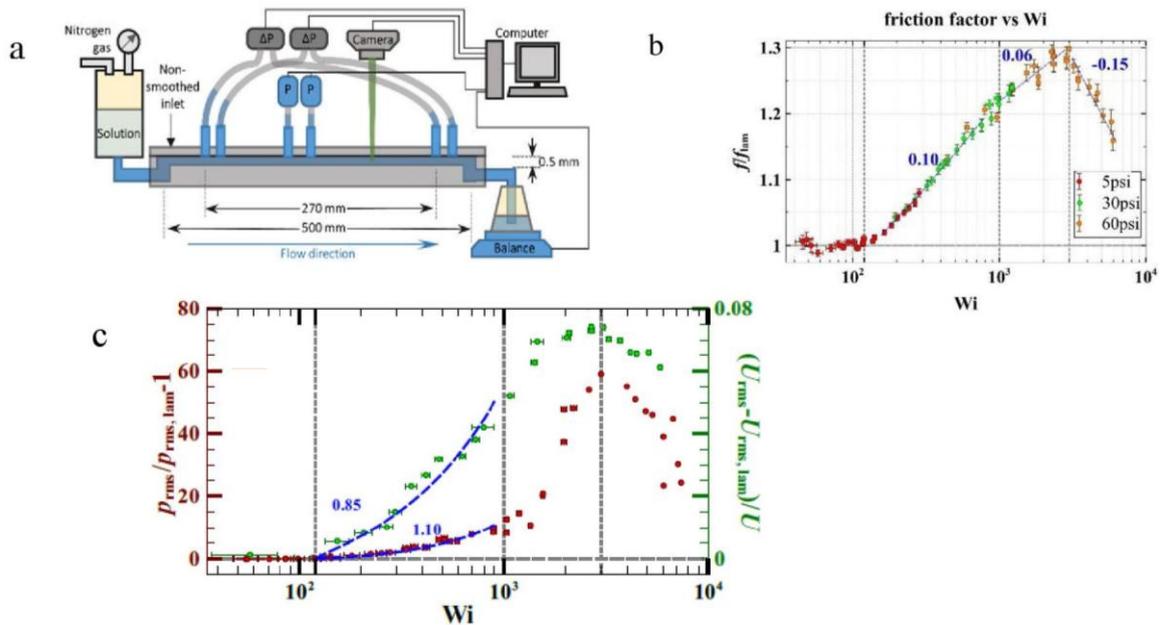

*Fig. 14.* (Color online) (a) Schematics setup. Six holes are used for pressure drops and fluctuations measurements. (b) $f/f_{lam}$ versus Wi. The pressure drop measurements carry out at two positions but with the same gap: from $L/h = 185$ to 725, and from 225 to 765, used. Three pressure sensors in 5 (red), 30 (green), and 60 (orange) psi are used. (c) Normalized rms pressure and velocity fluctuations versus Wi. Three vertical dash lines separate flow regimes.

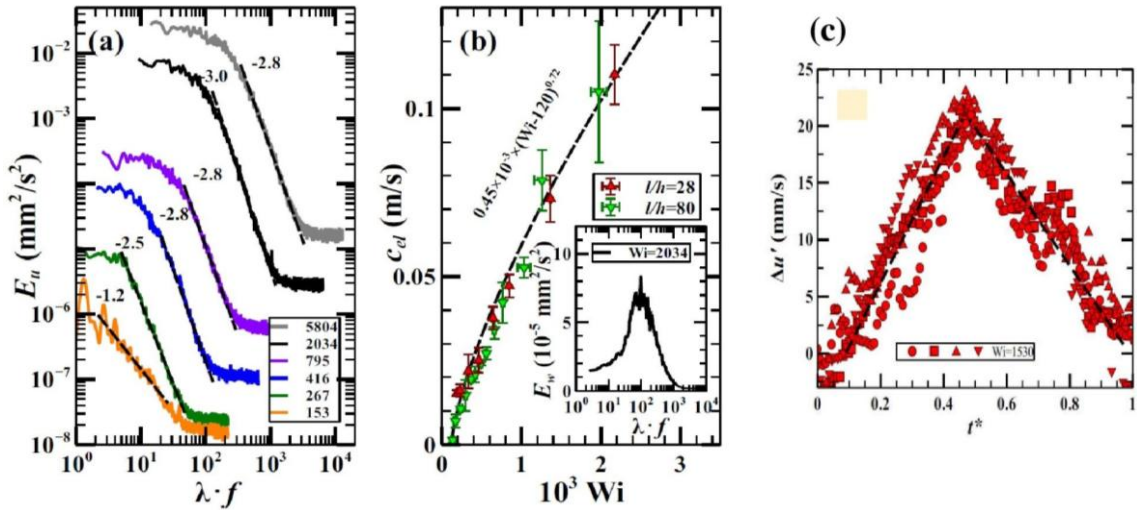

*Fig. 15.* (Color online) (a) Stream-wise velocity power spectra at channel center versus frequency, $\lambda f$, at $l/h = 330$. (b) Elastic wave speed versus Wi at $l/h = 28$ and 80. Inset: Elastic wave peak in span-wise velocity power spectrum at Wi = 2034 and $l/h = 330$. (c) Dynamics of streaks at $l/h = 330$ and Wi = 1530. Four runs of $\Delta u' = u'_2 - u'_1$, velocity difference across streak interface [64], for one cycle period $t^* = t f_{el}$.

To conclude, we return to the questions posed in the short preface at the beginning of the subsection 4.3. (i) We show that the strong prearranged perturbations are not necessary condition to get the elastic instability in inertia-less viscoelastic channel flows at Wi >> 1. Moreover, the scaling behaviors of the friction factor, normalized stream-wise velocity and pressure fluctuations with Wi, and stream-wise velocity power spectra in the transition regime identify the instability as a non-modal bifurcation. (ii) Three flow regimes are found above $\text{Wi}_c$: transition, ET, and DR, with streaks and elastic waves, similarly to those found in [64]. (iii) The velocity of stream-wise-propagating elastic waves has the same scaling exponent with $\text{Wi}-\text{Wi}_c$ and coefficient in various flow geometries as in [50, 64]. It suggests universality of the scaling relation. (iv) In three flow regimes only streaks, as CS, are observed, similar to the transition regime but in contrast to ET and DR, where both stream-wise rolls and streaks are found in a viscoelastic channel flow with strong prearranged perturbations at the inlet [64]. It occurs due to a lower intensity of the elastic waves in transition and, possibly, due to high sensitivity of CSs to initial conditions, similar to Newtonian channel flow [57]. (v) The remarkable finding is the existence of the streaks and elastic waves over the entire channel length up to $l/h = 980$, in a sharp contrast to a flow with the strong prearranged perturbations at the inlet [64], where CSs and elastic waves retain only from $l/h = 36$ up to $l/h \approx 200$.

## 5. Summary

To conclude, I summarize the findings and new observations presented in the review.

(i) The non-modal instability as the secondary bifurcation is observed in a flow between two obstacles hindering a viscoelastic channel flow and characterized for the first time. The following results of measurements in two mixing layers flow with parallel streamlines above the non-modal instability are used: the Wi dependence of the friction factor and time-averaged vorticity, the appearance of random flow with a velocity power spectrum with the exponent of its decay, corresponding to ET, and the elastic wave peak at low frequency.

(ii) At high El and Wi and Re << 1, the transition of ET to DR and down to relaminarization and elastic waves, which speed is quantified by scaling dependence on Wi and later found to be universal in other flow geometries, are discovered there.

(iii) Above the first non-modal instability in a viscoelastic straight channel flow with strong prearranged perturbations at the inlet, we observe CSs, rolls and streaks. They are synchronized by the elastic waves into cycling SSP in three flow regimes above the non-modal bifurcation: transition, ET, and DR. However, both the CSs and elastic waves are found in the limited range downstream from the perturbation source, and in the rest of the channel, one observes only velocity fluctuations until the outlet.

(iv) We verify that the strong prearranged perturbations at the channel inlet are not necessary condition to get an elastic instability in straight channel flows, and all channel flow arrangements exhibit the same properties of a non-normal mode instability. It is tested in two different channel arrangements: first, smoothed and tapered over a large length of the inlet with a small cavity at the top plate in the middle of the channel, and second, a straight channel with the unsmoothed inlet and six cavities at the top plate downstream the channel.

(v) It is remarkably found that the elastic waves propagate in two flow geometries in different directions: span-

wise in the first and stream-wise in the second. Moreover, the wave speed has the same scaling exponent for the $(\text{Wi} - \text{Wi}_c)$ dependencies in both cases but the coefficient *A* differs by almost three orders of magnitude and is smaller for the first case. The latter remains unexplained.

(vi) Due to extremely small velocity fluctuations in the first geometry, we are able to visualize the elastic waves and recover the linear dispersion relation. Furthermore, in the last two cases only streaks are shown up, probably, due to high sensitivity to perturbation spectra and lower intensity of the elastic waves.

(vii) Finally, the physical mechanism of relaminarization of ET is suggested and tested experimentally in a flow past an obstacle. We reveal quantitatively the correlation between the elastic wave intensity variations and friction factor, as well as qualitatively the correlation with wall-normal vorticity fluctuations by their imaging in ET and DR.

Finally, most of these observations resemble the properties of a Newtonian channel turbulent flow investigated theoretically and tested experimentally, in particular for the last two decades. Thus, it implies a similarity between Newtonian and viscoelastic channel flows, though these two flows are described by different equations: the Navier–Stokes versus elastic stress dynamics equations.

**Acknowledgement**

I am grateful to my collaborators on the recent developments in viscoelastic straight channel flows: A. Varshney, N. K. Jha, D.-Y. Li, M. V. Kumar, R. Shnapp, and Y. Li. This work was partially supported by grants from the Israel Science Foundation (ISF; grant \#882/15 and grant \#784/19) and the Binational USA-Israel Foundation (BSF; grant \#2016145).

___________________________

Новий напрямок та перспективи пружної нестійкості та турбулентності у різних в'язкопружних геометріях течії без інерції

Victor Steinberg

Представлено основні результати з пружно-керованими нестійкостями та пружною турбулентністю у в'язкопружних безінерційних течіях з криволінійними лініями струму. Описано теорію пружної турбулентності та передбачено пружні хвилі Re << 1 та Wi >> 1, швидкість яких залежить від пружної напруги, за аналогією з альвенівськими хвилями в магнітогідродинаміці, але на відміну від усіх інших, швидкість їх залежить від еластичності середовища. Оскільки при нульовій кривизні пропонований механізм пружної нестійкості в течіях з криволінійними лініями струму стає неефективним, доведено, що паралельні зсувні течії виявляються лінійно стійкими, подібно до ньютонівських паралельних зсувних течій. Проте лінійна стійкість паралельних зсувних течій не означає їх глобальної стабільності. Основна тема статті — останні розробки у розгляді паралельно-канального потоку. У такому потоці виявлено: пружна нестійкість, пружна турбулентність, пружні хвилі та зниження опору вниз аж до реламінаризації. Коротко обговорюються нормальні та ненормальні біфуркації у таких потоках.

Ключові слова: нестійкість, пружна турбулентність, в'язкопружні безінерційні потоки.